\documentclass[fleqn,usenatbib]{mnras}
\usepackage[utf8]{inputenc}
\usepackage[english]{babel}
\usepackage[T1]{fontenc}
\usepackage{ae,aecompl}
\usepackage{graphicx}
\usepackage{rotating} 
\usepackage{epsfig} 
\usepackage{amsmath}
\usepackage{amssymb}
\usepackage{euscript}
\usepackage{pdflscape}
\usepackage{xcolor}
\usepackage[normalem]{ulem}
\usepackage{url}
\definecolor{green2}{rgb}{0,0.6,0}
\definecolor{blue2}{rgb}{0.4,0.0,0.5}

\title[ZZ~Tau]{Orbital parameters and activity of ZZ~Tau -- a low mass young binary with circumbinary disc}

\author[A.~Belinski et al.]{
A.~Belinski,$^{1}$
M.~Burlak,$^{1}$
A.~Dodin,$^{1}$
N.~Emelyanov,$^{1}$
N.~Ikonnikova,$^{1}$
S.~Lamzin,$^1$\thanks{E-mail: lamzin@sai.msu.ru} 
\newauthor
B.~Safonov,$^1$
and
A.~Tatarnikov$^1$
\\
$^{1}$ Sternberg Astronomical Institute, Moscow M.V. Lomonosov State
University, Universitetskij pr., 13,  Moscow, 119992, Russia \\
}

\date{Accepted ~~~~~~~~~~~~~~~~~~~~~~~~~~~~~~~~. Received
~~~~~~~~~~~~~~~~~~~~~~~~~~~~~~; in original form }
\pubyear{2022}

\begin{document}
\label{firstpage}
\pagerange{\pageref{firstpage}--\pageref{lastpage}}

\maketitle

\begin{abstract}
We present the results of our new observations of the young binary ZZ~Tau
with a circumbinary disc. The system was found to consist of two coeval
(age $<2$~Myr) classical T~Tauri stars with the total mass $0.86 \pm
0.09$~M$_\odot$, orbital period $46.8 \pm 0.8$~yr, semimajor axis $88.2 \pm
2.1$~mas, eccentricity $0.58 \pm 0.02$ and the orbital inclination $123\fdg
8 \pm 1\fdg 0.$ The accretion rate of ZZ~Tau~A and ZZ~Tau~B are
approximately $7\times 10^{-10}$ and $2\times 10^{-10}$~M$_\odot$~yr$^{-1},$
respectively. No correlation was found between the long-term photometric
variability of ZZ~Tau and orbital position of its components.  The periodic
light variations with $P=4.171 \pm 0.002$~days was observed in the $BVRI$
bands presumably connected with an accretion (hot) spot on the surface of
the primary (ZZ~Tau~A).  At the same time no periodicity was observed in the
$U$ band nor in the emission line profile variations probably due to the
significant contribution of ZZ~Tau~B's emission, which dominates shortward
of $\lambda \approx 0.4$~\micron{}.  We argue that the extinction in the
direction to the primary is noticeably larger than that to the secondary. 
It appeared that the rotation axis of the primary is inclined to the
line of sight by $\approx 31 \degr \pm 4\degr.$ We concluded also that ZZ~Tau is the source of an CO molecular outflow, however, ZZ~Tau~IRS rather than ZZ~Tau is the source of the Herbig-Haro object HH393.

\end{abstract}

%
\section{Introduction}
\label{sect:introduct}

  In the last two decades, some young binary stars were discovered whose
orbital plane is noticeably inclined to the circumbinary (CB) disc
\citep{Smallwood-2021}. At first glance, it seems that this fact
contradicts the idea that the binary and its CB disc were formed from the 
same protostellar cloud. But the star formation in turbulent molecular cloud
may lead to chaotic accretion, as a result of which the misalignment between the 
disc and orbital plane of a binary may occur \citep{Monin-2007, Bate-2010,
Wurster-2019b}. Such systems can also originate from dynamical
encounters during star cluster formation \citep{Bate-2018}.

  \citet{Martin-2017} and \citet{Zanazzi-2018} have shown that if
a binary system has a noticeable eccentricity, then the initially small
$\Theta$ angle between the orbital plane and the CB disc either gradually
decreases to zero or increases to $90 \degr$ in a time-scale less than the
disc lifetime. It means that it is important to study young
binaries with different values of $\Theta$ to better understand the process
of star and planet formation.
  
 Here we consider a low mass young binary ZZ~Tau with a CB disc. 
The aim of the paper is to specify orbital parameters of the binary and to
discuss possible relation of the binary components activity to the
presence of a CB disc.

   ZZ~Tau is located within the dark cloud filament Barnard 18
\citep{Myers-1982}.  Its irregular variability in the range
$12\fm9-15\fm2$~pg was discovered by \citet{Reinmuth-1930}. 
\citet{Herbig-Rao-1972} included the star into the {\lq}Second Catalog of
Emission Line Stars of the Orion Population{\rq} and noted that {\lq}only
\ion{Ca}{ii} H,K lines have been observed in emission.{\rq} Later
\citet{Herbig-Bell-1988} detected the emission lines of \ion{He}{i} and H$\alpha$
(with an equivalent width of $EW_{{\rm H}\alpha}=15$~\AA{}), so ZZ~Tau was
classified as a T~Tauri (TTS) M3 type star in their Catalog (HBC~46).

 The {\it Gaia} parallax for ZZ~Tau ({\it Gaia} EDR3 id 147869784062378624)
is $7.46 \pm 0.14$~mas, which corresponds to a distance of $134 \pm 3$~pc
\citep{Gaia-16b, Gaia-2020}. But the respective astrometric solution
is not reliable with a renormalized unit weight error of 6.64,
so we will further use the distance $d=140$~pc as in the paper of
\citet{Schaefer-2014}.

  The 1991 lunar occultation observation of \citet{Simon-1995} revealed that
ZZ~Tau is a binary system with a projected separation of 29~mas. 
\citet{Schaefer-2014} found from the analysis of the previous
\citep{Schaefer-2003, Schaefer-2006} and new observations the following
orbital parameters of the system: period $P\approx 46$~yr, semimajor axis
$a\approx 12$~au, eccentricity $e\approx 0.57.$ It was also concluded that
the system consists of M2.5+M3.5 stars with a total mass of $M_{\rm
A}+M_{\rm B}\approx 0.8$~M$_\odot$ and an age of $< 3$~Myr in agreement with
later estimation of \citet{Zhang-2018}.

  The spectral energy distribution (SED) of ZZ~Tau\,A+B demonstrates an excess
emission longward of $\lambda \approx 4$~{\micron}\footnote{
Large veiling $r=0.7^{+0.3}_{-0.1}$ of the ZZ~Tau's spectrum at $\lambda =
4.7$~{\micron} found by \citet{Doppmann-2017} may be due to the fact that
the authors erroneously used in their analysis $T_{\rm eff}$ and $\log g$ of
ZZ~Tau~IRS -- another young star located $\approx 35$\arcsec{} to the south
from ZZ~Tau \citep{Gomez-1997}.  }
and prominent silicate features between 8 and 21~{\micron}
\citep{Sargent-2009}.  Analyzing these features \citet{Espaillat-2012}
concluded that at least one component of the binary has an optically thick
accretion disc -- see also Fig.\,7 and 8 in \citet{Furlan-11}.  It can be
assumed that the accretion of disc's matter is the reason for the observed
short wavelength excess emission in the optical
\citep{Herzeg-Hillebrandt-2014} and UV \citep{GALEX-2015} spectra of
ZZ~Tau\,A+B.  The accretion rate estimations $\dot M_{\rm acc}$ based on the
analysis of similar (spatially unresolved) spectra, vary between $2\times
10^{-10}$ \citep{Cieza-2012} and $1.3\times 10^{-9}$~M$_\odot$\,yr$^{-1}$
\citep{Valenti-1993}.

  The minimal distance between the components of the binary is 
$a\left( 1-e \right)<6$~au, so one can expect that the outer radii of their
circumstellar discs are $\lesssim 3$~au due to dynamical interaction
\citep{Artymowicz-Lubow-1994}. According to \citet{Espaillat-2012} it means
that the mass of these discs $M_{\rm disc}$ is $\lesssim 2\times
10^{-5}$~M$_\odot$ in agreement with the observed upper limit of $4\times
10^{-4}$~M$_\odot$ \citep{Andrews-2005, Andrews-2013}.  Probably it explains
why only weak continuum flux in 70 and 100~{\micron} bands and no line
emission (e.g.  [\ion{O}{i}] 63~{\micron}) was observed by
\citet{Howard-Hershel-2013}.

  As far as $M_{\rm disc}/\dot M_{\rm acc}$ time-scale is an order of
magnitude less than the age of the binary, the inner circumstellar discs
should be replenished in some way.  But the mass of the
circumbinary disc of ZZ~Tau\,A+B found by \citet{Akeson-2019} from {\it
ALMA} observations appears too small $(\log M_{\rm disc}/M_\odot =-4.24 \pm
0.11)$ to solve the problem. Then the question arises: whether the blue
excess emission of ZZ~Tau does refer to the accretion rather
than to the chromospheric activity.

  Neither forbidden lines nor other indications to outflow activity were found in the
optical \citep{Valenti-1993, Kenyon-1998, Herzeg-Hillebrandt-2014} and near IR
\citep{Folha-2001} spectra of ZZ~Tau\,A+B.  At the same time
\citet{Heyer-1987} found monopolar redshifted ${}^{12}$CO-outflow presumably
from ZZ~Tau, while \citet{Narayanan-2012} noted that the morphology of the
outflow admits the possibility that ZZ~Tau~IRS is the source of the outflow. 
Besides, \citet{Gomez-1997} detected an [\ion{S}{ii}] emission knot,
presumably a Herbig-Haro object (HH~393), and argued that ZZ~Tau~IRS is the
driving source for this outflow, but \citet{Bally-2012} suppose that
its source is ZZ~Tau.

  \citet{Schaefer-2014} wrote: {\lq}we suspect that the orbital
parameters{\rq} of the binary {\lq}will be revised substantially in the
future.{\rq} As we have seen, the physical parameters of ZZ~Tau's components
as well as the nature of their activity also need to be clarified.

 The rest of the paper is organized in the following way. Initially we
describe our observations (Section 2) and then present and discuss the
results in Section 3. Section 4 is devoted to the problems related to
circumstellar environment of the binary. Finally, we summarize the
conclusions.

%
\section{Observations}
 \label{sect:observation}

  The unresolved photometry of ZZ~Tau\,A+B was carried out with the 0.6-m
telescope of the Caucasian Mountain Observatory (CMO) of Sternberg
Astronomical Institute of Lomonosov Moscow State University (SAI MSU)
equipped with a CCD camera and a set of standard Bessel-Cousins $UBVR_{\rm c}I_{\rm c}$
filters \citep{Berdnikov2020}.  Two additional observations in the same
bands were performed with the 2.5-m telescope of the CMO equipped with a
mosaic CCD camera and a set of similar $UBVRI$ filters.  One can find a more
detailed description of the equipment, observational routine and data
reduction procedures in \citet{Dodin-19}. A single set of comparison stars
was used through all the observations, the $BVRI$ magnitudes for them were
adopted from {AAVSO}\footnote{https://www.aavso.org}.
The stars are: 000-BLB-648, 000-BLR-132, 000-BLR-134, 000-BLB-650,
000-BLR-079, 000-BLR-133. For the $U$ band, we adopted the comparison star
XEST 13-OM-054 from \citet{Audard-2007}.

   Results of our photometry are presented in
Table\,\ref{tab:RC-60-phot}\footnote{rJD abbreviation in the first column of
the Table means a reduced Julian Date $\rm rJD=JD-2\,450\,000$ and will be
used below as well.}.
%

%
\begin{table}
\renewcommand{\tabcolsep}{0.04cm}
 \caption{Optical photometry of ZZ~Tau\,A+B}
  \label{tab:RC-60-phot}  
\begin{tabular}{lllllllllll}
\hline
rJD & $U$ & $\sigma_{\rm U}$ & $B$ & $\sigma_{\rm B}$ & $V$ & $\sigma_{\rm V}$ & 
$R$ & $\sigma_{\rm R}$ & $I$ & $\sigma_{\rm I}$ \\
\hline
8382.51  &       &      & 15.78 & 0.10 & 14.37 & 0.02 & 13.10 & 0.07 & 11.19 & 0.10 \\ 
9104.508 & 15.88 & 0.05 &       &      &       &      &       &      &       &      \\
9187.550 & 15.35 & 0.05 & 15.91 & 0.07 & 14.37 & 0.03 & 12.93 & 0.06 & 11.18 & 0.10 \\
\hline
  \end{tabular} \\
The Table is available in its entirety in a machine-readable form in the
online journal. A portion is shown here for guidance regarding its
form and content.
\end{table}
%

  The unresolved near infrared (NIR) observations of ZZ~Tau were carried out
between 2015 December and 2021 November in the $JHK$ bands of the MKO photometric
system at the 2.5-m telescope of CMO SAI MSU equipped with the infrared
camera-spectrograph ASTRONIRCAM \citep{Nadjip-17}. The details of observations
and data reduction are described in \citet{Dodin-19}, the results are presented in Table\,\ref{tab:tab2}.

%
\begin{table}
\caption{NIR photometry of ZZ~Tau\,A+B} 
 \label{tab:tab2} 
 \begin{center}
\begin{tabular}{ccccccc}
\hline
rJD     & $J$ & $\sigma_J$ & $H$ & $\sigma_H$ & $K$ & $\sigma_K$ \\
\hline
9079.528 & 9.42 & 0.01  & 8.74 & 0.03 & 8.42 & 0.03 \\
9084.487 & 9.41 & 0.01  & 8.73 & 0.03 & 8.39 & 0.02 \\
9091.572 & 9.45 & 0.03  & 8.78 & 0.05 & 8.44 & 0.01 \\
9094.494 & 9.43 & 0.01  & 8.73 & 0.02 & 8.46 & 0.02 \\
9551.258 & 9.54 & 0.07  & 8.76 & 0.02 & 8.45 & 0.03 \\
\hline
\end{tabular}
\end{center}
\end{table}
%

  We also carried out polarimetric observations of ZZ~Tau\,A+B in the
$I_{\rm c}$ band with the SPeckle Polarimeter (SPP) of the 2.5-m telescope
of SAI MSU \citep{Safonov-17}. The details of observations and data reduction
are described in \citet{Dodin-19}, the results are presented in
Table~\ref{tab:polarim}.

\begin{table}
\renewcommand{\tabcolsep}{0.15cm}
\caption{Optical polarimetry of ZZ~Tau\,A+B in $I$ band.}
 \label{tab:polarim} 
\begin{tabular}{ccccc} 
\hline 
rJD  & $p$ & $\sigma_{\rm p}$ & $\theta$ & $\sigma_{\theta}$ \\
     & \%  & \%               & $\degr$  & $\degr$           \\ 
\hline 
7818.25 & 0.27 & 0.03  & 118.6 & 3.4 \\
7820.22 & 0.34 & 0.03  & 118.2 & 2.6 \\
9514.52 & 0.21 & 0.07  & 116.9 & 8.9 \\
9517.43 & 0.28 & 0.04  & 122.9 & 3.8 \\  
\hline 
\end{tabular}\\ 
Col.\,1: Date of observation;
Col.\,2--3: the polarization degree and its error;
Col.\,4--5: the polarization angle and its error;
\end{table}
%

 SPP was also used for speckle interferometric observations of ZZ~Tau. 
Observations were carried out on 2019, December, 17 and 2021, October, 27 in
the passband centered at $0.9~\micron,$ which is close to (but does not
coincide with) the $I_{\rm c}$ band. The difference of the binary
components magnitues at these moments were $0.71 \pm 0.10$ and $0.65 \pm
0.10.$ Respective values of angular separation $\rho$ and position angle
$PA$ are presented in two last rows of Table~\ref{tab:astrometry}.

  For our study we have additionally used high and medium resolution optical
spectra of ZZ~Tau taken from the archives of the following telescopes: Keck
(HIRES spectrograph, spectral resolution $R=\lambda/\Delta \lambda \approx
48\,000,$ PI D.~Scott),\footnote{
https://koa.ipac.caltech.edu/cgi-bin/KOA/nph-KOAlogin}
Canada-France-Hawaii (ESPaDOnS,
$R\approx 65\,000,$ PI L.~Cieza),\footnote{
http://www.cadc-ccda.hia-iha.nrc-cnrc.gc.ca/en/}
and Very Large Telescope:\footnote{http://archive.eso.org/scienceportal/home}
spectrographs UVES ($R\approx 42\,000,$ PI E.~Moraux) as well as MUSE
($R\approx 2500,$ PI S.~Haeffert). The spectra obtained with HIRES,
ESPaDOnS and UVES spectrographs are combined spectra of both components of
ZZ~Tau, whereas MUSE resolves the spectra of each component of the binary. 
Some additional information about these spectra and their abbreviated names
that will be used below are presented in Table~\ref{tab:spectra}.

%
\begin{table}
\renewcommand{\tabcolsep}{0.15cm}
 \caption{Spectra of ZZ~Tau analysed in this paper} 
  \label{tab:spectra}
  \begin{center}
\begin{tabular}{lllll}
\hline
Object     & Date, UT            & Name   & Band, nm  \\
\hline
ZZ~Tau\,A$+$B &  2008-Dec-04, 14:02 & HIRES  & 445-890   \\
              &  2009-Sep-30, 01:26 & ESP-1  & 370-1048  \\
              &  2010-Jul-30, 03:22 & ESP-2  & 370-1048  \\
              &  2010-Dec-17, 01:26 & ESP-3  & 370-1048  \\
              &  2018-Nov-27, 07:00 & UVES-1 & 473-684   \\
              &  2018-Dec-07, 06:19 & UVES-2 & 473-684   \\
              &  2018-Dec-24, 01:58 & UVES-3 & 473-684   \\
ZZ~Tau A,\,B  &  2020-Jan-14, 02:07 & MUSE   & 475-935   \\
\hline
  \end{tabular}
 \end{center}
\end{table}
%

\begin{figure}
 \begin{center}
\includegraphics[scale=0.40]{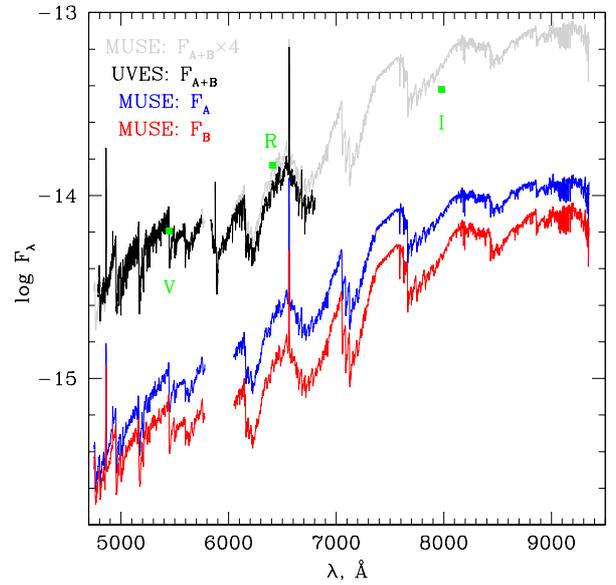}
 \end{center}
  \caption{Observed MUSE spectra of ZZ~Tau~A (blue) and ZZ~Tau~B (red).  The
sum of component's spectra increased by 4 times (grey) is shown to reconcile
with our photometric measurements in the $V,$ $R$ and $I$ bands
(green squares) as well as with the UVES-1 spectrum of ZZ~Tau~A+B degraded to
spectral resolution R=2500 (black).  The flux is in units erg
s$^{-1}$\,cm$^{-2}$\,\AA$^{-1}.$ See text for details.}
 \label{fig:MUSE-vs-CMO}
\end{figure}

  All the spectra are marked as a {\lq}scientific grade{\rq} in the
archives, so we did not process them additionally.  The spectra of ZZ~Tau
observed with the HIRES and ESPaDOnS spectrographs are normalised to the
continuum level, whereas the spectra observed with the UVES and MUSE
spectrographs are flux calibrated.  However, the sum of {\it absolute}
fluxes of ZZ~Tau\,A and B components in the MUSE spectrum appeared $\approx
4$ times less than the fluxes corresponding to our $VRI$ photometric data --
see Fig.~\ref{fig:MUSE-vs-CMO}.  In contrast, all the three UVES spectra,
which by the way coincide with each other (but see
Sect.~\ref{subsect:activity}), are in agreement with the results of our $V$
and $R$ photometry.  As can be seen from the figure, the UVES spectra also
coincide with the 4 times increased combined MUSE spectrum up to $\approx
0.62$~\micron{}, i.e.  almost in the entire $V$ band.

  We did not find the source of wrong flux calibration of the {MUSE}
spectra, and did not use information about the absolute fluxes of ZZ~Tau
from these spectra in our study.  At the same time we have no reason to
doubt that the flux ratio of ZZ~Tau's components in the MUSE spectrum
shortward of $\sim 0.6$~\micron{} is correct.

%
%

%
\section{Results and discussion}
 \label{sect:results}
%

\subsection{The Orbit}
 \label{subsect:orbit}
%
%
\begin{table}
\caption{Astrometric observations of ZZ~Tau}
 \label{tab:astrometry}
  \begin{tabular}{cccccc}
\hline
Date      & $\rho$ & $\sigma_{\rm \rho}$ & PA      & $\sigma_{\rm PA}$ & Ref. \\
yr        & mas    &  mas                & $\degr$ & $\degr$           &      \\
\hline
1995.081  & 36.0   & 7.6    & 187.0   & 11    &  1 \\
1996.124  & 36.0   & 7.6    & 177.0   & 10    &  1 \\
1997.704  & 50.6   & 5.4    & 129.2   & 6.0   &  1 \\
1998.214  & 55.0   & 5.4    & 125.0   & 6.0   &  1 \\
1999.621  & 57.9   & 1.3    & 112.9   & 1.8   &  1 \\
2000.564  & 59.5   & 1.2    & 106.2   & 1.8   &  1 \\
2000.690  & 58.2   & 1.2    & 105.7   & 1.8   &  1 \\
2001.646  & 62.5   & 1.1    &  98.9   & 1.8   &  1 \\
2002.748  & 60.6   & 1.3    &  91.1   & 1.7   &  1 \\
2002.929  & 61.2   & 1.5    &  88.8   & 1.5   &  1 \\
2003.723  & 62.8   & 1.5    &  87.6   & 1.4   &  2 \\
2004.982  & 61.32  & 1.35   &  74.02  & 1.26  &  3 \\
2005.849  & 61.7   & 1.5    &  67.2   & 1.4   &  2 \\
2006.963  & 66.31  & 1.21   &  61.39  & 1.05  &  3 \\
2008.045  & 66.31  & 0.49   &  54.09  & 0.42  &  3 \\
2008.962  & 67.75  & 0.55   &  48.80  & 0.47  &  3 \\
2011.065  & 73.02  & 1.05   &  38.39  & 0.82  &  3 \\
2011.780  & 74.10  & 0.41   &  33.89  & 0.32  &  3 \\
2013.074  & 76.77  & 1.56   &  23.74  & 1.16  &  3 \\
2014.068  & 79.52  & 0.63   &  23.37  & 0.46  &  3 \\
2019.961  & 93.5   & 2      &   0.5   & 1.0   &  4 \\
2021.819  & 99.0   & 1      & 358.1   & 0.8   &  4 \\
\hline
  \end{tabular} \\
1 -- \cite{Schaefer-2003}, 2 -- \cite{Schaefer-2006}, \\
3 -- \cite{Schaefer-2014}, 4 -- this work. Additional \\
information from 1991.592 lunar occultation:$\rho=29$~mas to \\
PA$=244\degr$  \citep{Simon-1995}.
\end{table}
%

%
\begin{table*}
\renewcommand{\tabcolsep}{0.18cm}
\caption{Orbital parameters of ZZ~Tau} 
 \label{tab:orbit}
 \begin{center}
\begin{tabular}{ccccccccc}
\hline
Source  & $P$ & $T_0$ & $e$ & $a$ & $i$ & $\Omega$ & $\omega$  & $M_{\rm A}+M_{\rm B}$ \\
        & yr  & yr    &     & mas  & $\degr$ & $\degr$ & $\degr$ & M$_\odot$ \\
\hline
\citet{Schaefer-2014} & $46.2 \pm 2.8$ & $1995.01 \pm 0.28$ & $0.57 \pm 0.03$ & $86.4 \pm 4.5$
& $124.6 \pm 2.0$ & $137.0 \pm 2.3$ & $294.9 \pm 2.3$ & $0.83 \pm 0.16$ \\
This work         & $46.8 \pm 0.8$ & $1995.07 \pm 0.13$ & $0.58 \pm 0.02$ & $88.2 \pm 2.1$  
& $123.8 \pm 1.0$ & $136.7 \pm 1.3$ & $294.9 \pm 0.9$ & $0.86 \pm 0.09$ \\
Mirror orbit      & $46.8 \pm 0.8$ & $1995.07 \pm 0.13$ & $0.58 \pm 0.02$ & $88.2 \pm 2.1$  
& $123.8 \pm 1.0$ & $316.7 \pm 1.3$ & $114.9 \pm 0.9$ & $0.86 \pm 0.09$ \\
\hline
\end{tabular}
\end{center}
\end{table*}
%

  To calculate the orbit of ZZ~Tau we used the same observational data as
\citet{Schaefer-2014} -- see Table~\ref{tab:astrometry} -- including
the constraints implied by the 1991.6 lunar occultation observation of
\citet{Simon-1995} and added the results of our measurements.  The orbit was
determined by the differential refinement of orbital parameters using the
least squares method (see \citet{Emelya-2020} for details). Weights were assigned to the observations according to the errors given by the observers. The observed minus calculated positions were found for each observation. For these deviations the RMS of the post fit to the complete data set are equal to 3.1 mas for an unweighted observation and to 0.98 mas for a weighted one.

  As can be seen from comparing the first and the second rows of
Table~\ref{tab:orbit}, our values of orbital period $P,$ time of periastre
passage $T_0,$ eccentricity $e,$ (angular) semimajor axis $a,$ orbital
inclination $i,$ position angle of ascending node $\Omega,$ argument of
periastre $\omega$ and total mass of the components $M_{\rm A}+M_{\rm B}$
are very close to the respective values found by
\citet{Schaefer-2014} but more accurate. As far as ZZ~Tau is a visual binary, 
$\omega$ is given for the secondary. With the adopted distance to the
binary 140~pc its semimajor axis in linear scale is $a\approx 12.3$~au.

  If using only astrometric data, it is not possible to determine which node
of the orbit is ascending and which is descending.  In other words the
projection of the orbit to the celestial sphere will be the same as in the
case of a{\lq}mirror{\rq} orbit, when the position angle of the ascending
node $\Omega$ and the argument of periastre $\omega$ presented in the
Table~\ref{tab:orbit} are simultaneously replaced by $\Omega+180\degr$ and
$\omega-180\degr,$ respectively -- see the third row in the table.  This
ambiguity can be eliminated by the fact that the observed radial velocity of
the companion relative to the primary $\Delta V_{\rm r}=V_{\rm r}^{\rm B} -
V_{\rm r}^{\rm A}$ in both cases takes the same absolute value in each point
of the orbit, but with opposite signs.\footnote{We used a Cartesian coordinate system with the origin in the primary, the $XY$ plane coinciding with the celestial sphere and the
$X,$ $Y$ and $Z$ axes directed to the north pole $N,$
east $E$ and from the Earth to the star, respectively.}

  Indeed, if $v$ is a true anomaly, i.e. the angle in the orbital plane
between the directions to the periastre and the companion viewed from the
primary, then it can be shown that \citep{Emelya-2020}:
\begin{equation}
\Delta V_{\rm r} = V_0\,
\left[ \, \cos \left( v+\omega \right) + e\, \cos \, \omega \, \right] \, , 
  \label{eq:delta-Vr-1}
\end{equation}
where, according to Table~\ref{tab:orbit},
\begin{equation}
V_0 = \sqrt{
{ G\left( M_{\rm A}+M_{\rm B} \right) \over a\left( 1-e^2 \right) }
} \times \sin i \approx 7.94\,\, \mbox{km s}^{-1}\, .
 \label{eq:delta-Vr-2}
\end{equation}
Thus, $\Delta V_{\rm r}$ varies between approximately $-6.0$ and
$+9.9$~km~s$^{-1}$, if $\omega=294\fdg 9$, or between $-9.9$ and
$+6.0$~km~s$^{-1}$ in the case of the mirror orbit $(\omega=114\fdg 9).$

  Radial velocity measurements based on spatially resolved NIR spectra of
ZZ~Tau's components (obtained with NIRSPAO at the Keck Observatory) were
kindly presented to us by L.Prato on our request.\footnote{
 See also http://jumar.lowell.edu/BinaryStars/}
The results of these measurements (with errors of $< 1$~km~s$^{-1}$)
are presented in Table~\ref{tab:Prato} along with the theoretical $\Delta V_{\rm
r}$ values calculated from Eq.(\ref{eq:delta-Vr-1}) for the moments of
respective observations, so that a {\lq}plus{\rq} sign of these values
corresponds to the values of $\omega$ and $\Omega$ from the second row of
Table~\ref{tab:orbit}, whereas a {\lq}minus{\rq} sign corresponds to the
{\lq}mirror{\rq} orbit.  As far as the observed $\Delta V_{\rm r}$ values are
negative the mirror orbit looks preferable, but see
Sect.~\ref{subsect:activity}.

%
\begin{table}
 \caption{Radial velocity (km\,s$^{-1})$ of ZZ~Tau's components} 
  \label{tab:Prato} 
  \begin{center}
\begin{tabular}{ccc}
\hline
Component & 2009, Dec 06 & 2010, Dec 10 \\
\hline
ZZ~Tau\,A &  $+16.20$       & $+15.77$        \\
ZZ~Tau\,B &  $+15.35$       & $+13.64$        \\
$\Delta V_{\rm r},$ obs & $-0.85$ & $-2.13$ \\
$\Delta V_{\rm r},$ theor  & $\pm 1.69$ & $\pm 1.27$ \\
\hline
  \end{tabular}
 \end{center}
\end{table}
%

\begin{figure}
 \begin{center}
\includegraphics[scale=0.6]{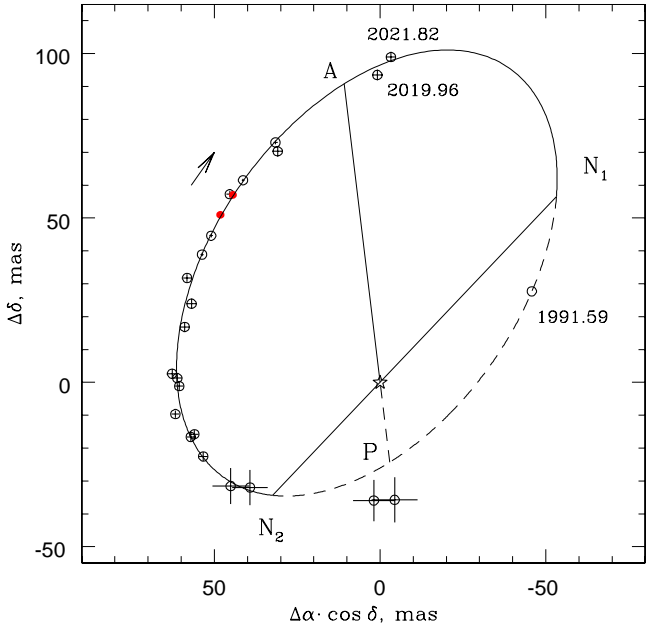}
 \end{center}
  \caption{Variations of relative position of components in ZZ~Tau. Open 
circles correspond to observation used to calculate the orbit. $A,$ $P$
letters mark the position of apoastron and periastron, respectively, and
${\rm N}_1-{\rm N}_2$ is the line of nodes, so that N$_1$ is the ascending
node.  The origin of the coordinate system is in the A component, the north
is up and the east is to the left. Two red points correspond to the moments
when the radial velocity of the components (presented in Table
\ref{tab:Prato}) was measured -- see text for details.
}
 \label{fig:orbit}
\end{figure}

  The comparison of observational data with the mirror orbit is shown in
Fig.~\ref{fig:orbit}, so that the part of the orbit that is closer to the
Earth than ZZ~Tau\,A $(z<0)$ is shown with a solid line, so the N$_1$ point
corresponds to the ascending node.  The weighted standard deviation $\sigma$
of theoretical orbit from observational data is $\approx 1.0$~mas.  We
believe that our orbit reproduces the observations fairly well, so the
discrepancy between the theoretical and observed $\Delta V_{\rm r}$ values
in Table \ref{tab:Prato} requires an explanation.  We will discuss a possible
reason for this disagreement in Sect.~\ref{subsect:activity}.

  It follows from our solution that the minimal and maximum distances
between the binary components are $r_{\rm min}= a(1-e) \approx 5.2$ and
$r_{\rm max}= a(1+e) \approx 19.5$~au, respectively, and the mass of the
system is $M_{\rm A}+M_{\rm B}= 0.86 \pm 0.09$~M$_\odot.$

%
%


\subsection{Photometry}
 \label{subsect:photometry}
%

%
\begin{figure}
 \begin{center}
\includegraphics[scale=0.45]{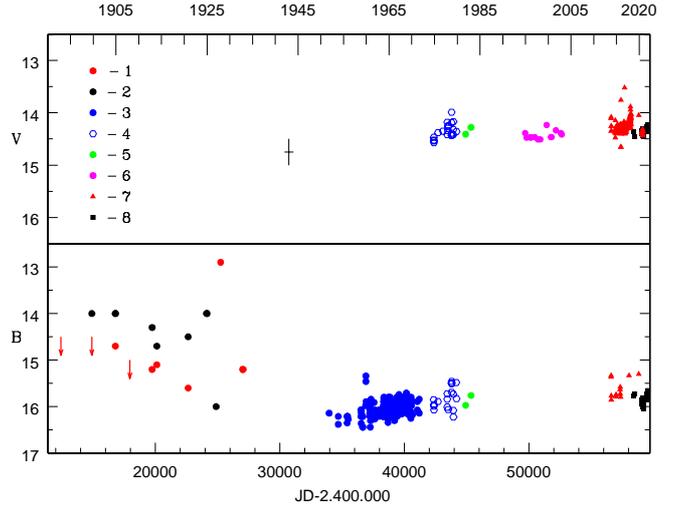}
 \end{center}
  \caption{The historical light curve of ZZ~Tau\,AB in the $V$ (upper panel) and $B$
(low panel) bands.  Simbols of different colour and/or shape correspond to
different data sources: 1 -- \citet{Reinmuth-1930}, 2 -- \citet{Himpel-1944},
3 -- \citet{Malyshev-1972}, 4 -- \citet{Nurmanova-1983}, 
5 -- \citet{Rydgren-83}, 6 -- \citet{Schaefer-2003}, 7 -- AAVSO, 8 -- 
this work.  }
 \label{fig:lcHist}
\end{figure}
%

  The historical light curve of ZZ~Tau\,A+B, shown in Fig.~\ref{fig:lcHist}, was
constructed from our $B$ and $V$ data (see Table~\ref{tab:RC-60-phot}) and
the data adopted from the literature \citep{Reinmuth-1930, Malyshev-1972, 
Kirillova-1963, Rydgren-83, Nurmanova-1983, Schaefer-2003}. 
The cross in the upper panel of the figure corresponds to the information from
\citet{Himpel-1944} that the 5 visual brightness estimations made for ZZ~Tau between 
JD~2\,430\,360.5 and 2\,430\,725.0 were in the range $14\fm 75-15\fm0.$

 As can be seen from the figure the average brightness of ZZ~Tau in the $B$ band
smoothly decreased till the beginning of 1950s and then began to
increase, but the amplitude of variability significantly decreased.  We can
not say anything about the behaviour of the star in the $V$ band before 1942 due
to the absence of data, but note that the {\it average} $V$ magnitude of
ZZ~Tau was nearly constant in the second half of the 20th century: e.g., compare
$14\fm 34$ from \citet{Herbig-Bell-1988}, $14\fm 35$ from \citet{KH-1995}
with our $14\fm 38$ in 2020 (see Fig.~\ref{fig:period}).

%
\begin{figure}
 \begin{center}
\includegraphics[scale=0.40]{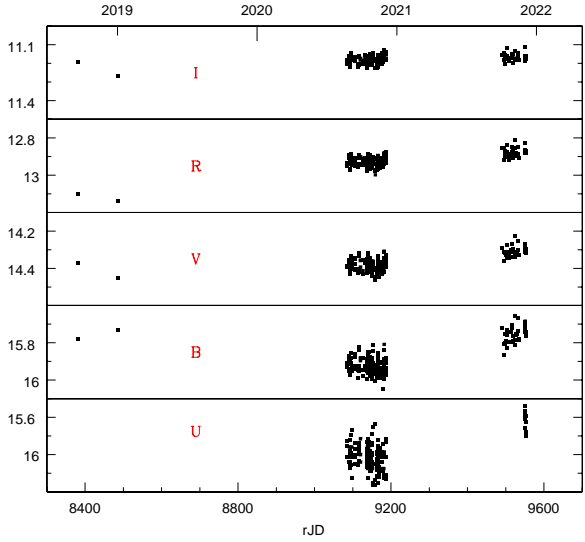}
 \end{center}
  \caption{The $UBVRI$ light curves of ZZ~Tau~A+B based on our data
from Table~\ref{tab:RC-60-phot} only. Note that the binary is brighter
in the second half of the 2021 season than in 2020.}
 \label{fig:lcCMO}
\end{figure}
%

  A more detailed light curve based on our observations only is presented in
Fig.\ref{fig:lcCMO}.  It can be seen that the binary becomes
brighter in all the bands in 2021 than in 2020, and to a greater extent in the
short-wavelength region. We will discuss this feature in
Sect.\ref{subsect:activity}.

%
\begin{figure}
 \begin{center}
\includegraphics[scale=0.45]{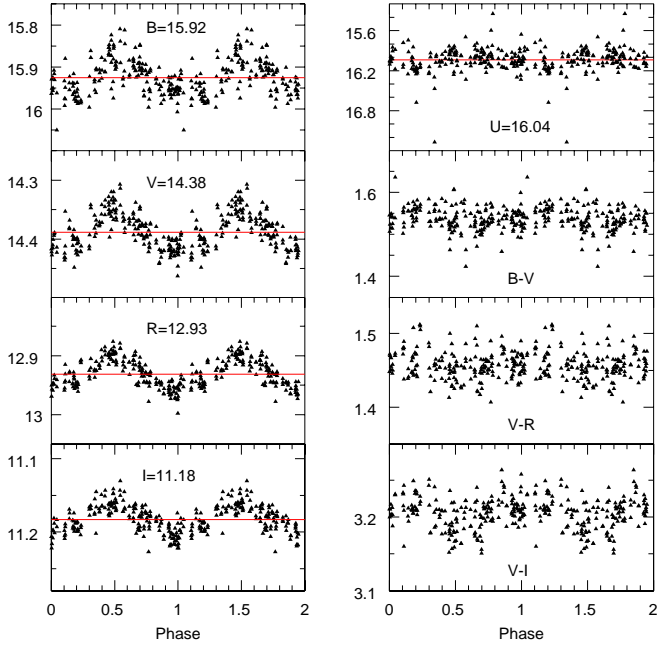}
 \end{center}
  \caption{The $UBVRI$ and colours phase curves of ZZ~Tau corresponding to 
the period $P\approx 4{\fd}17$, based on the data from Tabl.~\ref{tab:RC-60-phot}. 
The average brightness of the binary in each band is shown by a red
line in respective panels. Note that periodicity is clearly seen in the
panels of the left column in contrast to those of the right one. }
 \label{fig:period}
\end{figure}
%

   \citet{Rodriguez-2017} found periodical brightness variations of ZZ~Tau\,A+B
from ground based observations and attributed them to rotation modulation
with the period $P_{\rm rot} = 1{\fd}311.$ \citet{Rebull-2020} did not confirm
this period from the analysis of the space K2 data, but found another period
$P_{\rm rot} = 4{\fd}1609.$ Based on our data we found the following values 
$4{\fd}1688,$ $4{\fd}1702$ and $4{\fd}1739,$ for the $V,$ $R$ and $I$ bands, 
respectively. The average value $P{\rm rot}=4{\fd}1710\pm0{\fd}0022$ is 
close to the result obtained by \citet{Rebull-2020}. The shorter period $1\fd311$ is
the one day alias of the real one.

  The phase curves in different optical bands as well as $B-V,$ $V-R$ and $V-I$
colours are shown in Fig.~\ref{fig:period}. It can be seen from the
left column of the figure, that the less the effective wavelength of the
photometric band the more blurred becomes the phase curve, and in the $U$ band
the periodicity disappears. We will discuss this feature in
Sect.\ref{subsect:activity}. The periodicity is only 
marginally seen in the colours, probably because the amplitude
of the variability is comparable to the errors of measurements.

   As follows from Table~\ref{tab:tab2} the $J$ and $H$ magnitudes of
ZZ~Tau\,A+B were constant during our observations within errors of
measurements $(\bar J =9\fm42,$ $\bar H=8\fm74),$ and small variability with an
amplitude of $\approx 0\fm03$ can be suspected in the $K$ band $(\bar
K=8\fm42).$ One can compare these results with previous NIR observations:
for example $J=9.52\pm 0.02,$ $H=8.78 \pm 0.02$, $K=8.54\pm 0.02$
\citep{Rydgren-83} or $J=9.46\pm 0.02,$ $H=8.72\pm 0.02,$ $K=8.51\pm 0.02$
\citep{Whitney-1997}.


\subsection{Parameters of ZZ Tau's components}
 \label{subsect:TLAv}

  Based on spatially resolved IR spectra of the
binary \citet{Schaefer-2014} found the effective temperature $T_{\rm eff},$ 
extinction $A_{\rm V}$ and bolometric luminosity $L_{\rm bol}$ for each component.  
We have no such spectra, so we have to use the results of these
authors hereinafter: $T_{\rm eff}^{\rm A}=3488 \pm 145$~K, $L_{\rm bol}^{\rm A}=0.411
\pm 0.038$~L$_\odot$ and $T_{\rm eff}^{\rm B}=3343 \pm 145$~K, $L_{\rm
bol}^{\rm B}=0.241 \pm 0.055$~L$_\odot.$ Then the radii of the components
are $R^{\rm A} \approx 1.76 \pm 0.17$ and $R^{\rm B} \approx 1.47 \pm
0.21$~R$_\odot.$

To estimate the age and mass of the binary components
\citet{Schaefer-2014} used theoretical tracks and isochrones of
\citet{Baraffe+98, Siess-2000, Dotter-2008, Tognelli-2011}.  It was found
that the age of the system is $<3$~Myr and {\lq}for all sets of tracks, the
sum of the evolutionary masses of the components agrees at the $1\sigma$
level with the total dynamical mass{\rq} $M_{\rm A+B}=0.83 \pm
0.16$~M$_\odot.$

%
\begin{figure}
 \begin{center}
\includegraphics[scale=0.4]{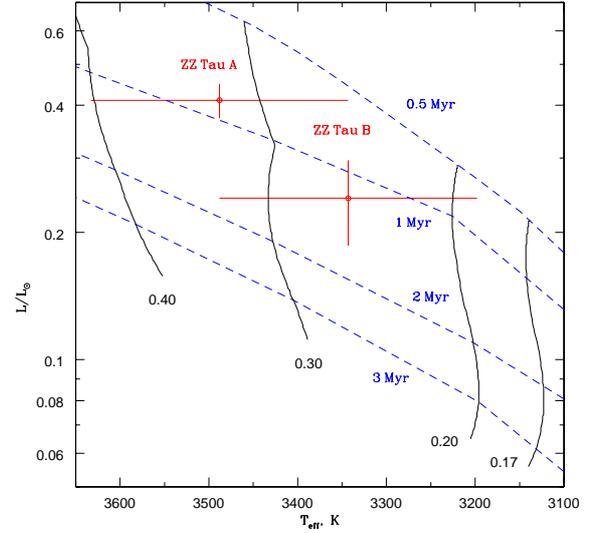}
 \end{center}
  \caption{The position of ZZ~Tau\,A and ZZ~Tau\,B on Hertzsprung-Russel
diagram. The evolutionary tracks (solid lines) and isochrones (dashed lines)
adopted from \citet{Baraffe+2015} are also plotted.  }
 \label{fig:HR-diagramm}
\end{figure}
%

We also plot the positions of ZZ~Tau's components in the
Hertzsprung-Russel diagram (see Fig.~\ref{fig:HR-diagramm}), but use more
recent theoretical calculations of \citet{Baraffe+2015}.  It follows from
the plot that the mass of the primary is $M_{\rm A}=0.33 \pm
0.08$~M$_\odot$ and that of the secondary is $M_{\rm B}=0.26\pm
0.06$~M$_\odot.$ Thus, the total mass of the binary $M_{\rm A+B}$ is $0.59\pm
0.10$~M$_\odot,$ that is somewhat less than the mass derived from our orbital
solution: $0.86 \pm 0.09$~M$_\odot.$ This discrepancy will disappear if the
actual distance $d$ to the star is 5-10 \% less than the accepted value of $140$~pc, because the dynamical mass is proportional to $d^3.$ Cold and hot spots on the surfaces of ZZ~Tau's components (see the next section) can also alter the estimation of their $T_{\rm eff}$ and `the inferred stellar masses from stellar evolutionary models' \citep{Flores-2022}. As follows from the figure the age of the system is between 0.5 and 2~Myr.

\begin{figure}
 \begin{center}
\includegraphics[scale=0.4]{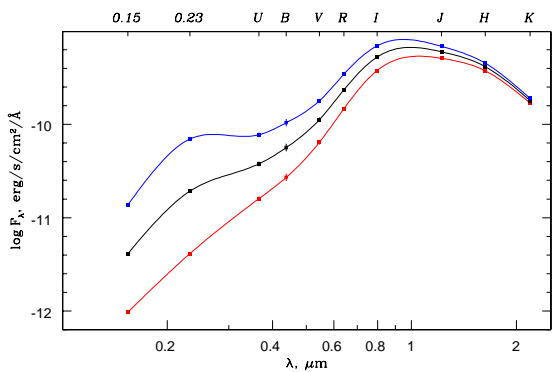}
 \end{center}
  \caption{The spectral energy distribution of ZZ~Tau~A+B based on our
and {GALEX} \citep{GALEX-2015} broad band photometric data (squares):
observed (red) and dereddened with $A_{\rm V}=0.6$ (black) and
1.1 (blue), respectively. The curves are spline interpolations of the data and 
are drawn for clarity. See text for details.}
 \label{fig:sed-AB}
\end{figure}

The luminocities and radii of the components depend on the extinction
adopted by \citet{Schaefer-2014}: $A_{\rm V}^{\rm A}=1.49 \pm 0.34,$ $A_{\rm
V}^{\rm B}=1.24 \pm 0.87.$ At the same time $A_{\rm V}^{\rm A+B}$ found by
\citet{Herzeg-Hillebrandt-2014} from spatially unresolved spectra of
ZZ~Tau~A+B is significantly less: $0.55 \pm 0.1.$ Spectral energy
distribution of ZZ~Tau~A+B based on our and {GALEX} \citep{GALEX-2015} broad
band photometric data (Fig.\ref{fig:sed-AB}) also indicate that the
situation with extinction is not clear.

%
%

  To convert the $UBVRIJHK$ and $FUV,$ $NUV$ magnitudes to monochromatic
fluxes $F_{\rm \lambda}$ we used the constants from the papers of
\citet{Bessell-98} and \citet{Morrissey-2007}, respectively.  Then we
dereddened observed fluxes assuming the standard extinction law with $R_{\rm
V}=3.1$ \citep{Cardelli+1989}.  As follows from the figure the $F_{\rm
\lambda}(\lambda)$ dependence has a maximum shortward of the $U$ band for
$A_{\rm V} > 1$. But if the observed UV emission of ZZ~Tau~A+B is
connected with an accretion spot (see the next section), then the accretion
shock theory predicts \citep{Lamzin-1995,Calvet-Gullbring-98,Dodin-18} that
the flux shortward of $\lambda=0.36$~\micron{} should decrease with decreasing
wavelength.  For this reason, we believe that the appearance of the maximum in
the $F_{\rm \lambda}(\lambda)$ curve is the result of excess de-reddening
which occurs due to the presence of a local maximum (bump) in the extinction
curve $A_\lambda(\lambda)$ at $\lambda \approx 0.22$~\micron{}.

  If so, then $A_{\rm V}$ should be $<1.0,$ however this statement is based
on the assumption that the continuum rather than line emission (e.g. 
\ion{C}{iv}~1550 and \ion{Mg}{ii}~2800 doublets) is the main sources of
radiation in FUV and NUV {GALEX}'s spectral bands.  Future spectral
observations of ZZ~Tau in the UV band -- e.g.  with {WSO-UV}
\citep{Boyarchuk-2016} -- will make it possible to find out if this is the
case.  X-ray observations of the binary will also help to to clarify $A_{\rm
V}.$

  According to \citet{Schaefer-2014} $A_{\rm V}^{\rm A} > A_{\rm V}^{\rm B}$
but the difference of these values is less than $1\sigma,$ so it well can be
that the extinction is the same for the primary and the secondary.  Note
however that shortward of $\lambda \approx 0.6$~nm the flux of ZZ~Tau~A
decreases faster than that of ZZ~Tau~B with the wavelength decreasing (see
Fig.\ref{fig:MUSE-vs-CMO}), although it should be the opposite, since the
effective temperature of the companion is less than that of the primary. A
smaller extinction in the direction to ZZ~Tau~B could eliminate this
discrepancy.


\subsection{The nature of the component's activity}
 \label{subsect:activity}

\begin{figure}
 \begin{center}
\includegraphics[scale=0.5]{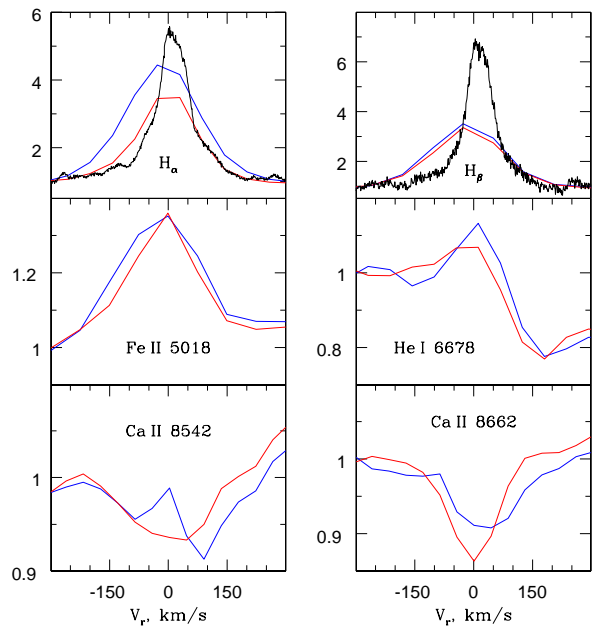}
 \end{center}
  \caption{The relative intensities of the most prominent lines in the MUSE
spectrum of ZZ~Tau\,A and ZZ~Tau\,B (blue and red curves, respectively). The 
UVES-1 spectrum of ZZ~Tau~A+B is shown in upper panels for comparison (black
curves). The profiles in each panel are normalised to the nearby continuum level. 
The heliocentric radial velocity is plotted along the horizontal axis.
}
 \label{fig:muse-emiss-lines} 
\end{figure}

\begin{figure}
 \begin{center}
\includegraphics[scale=0.4]{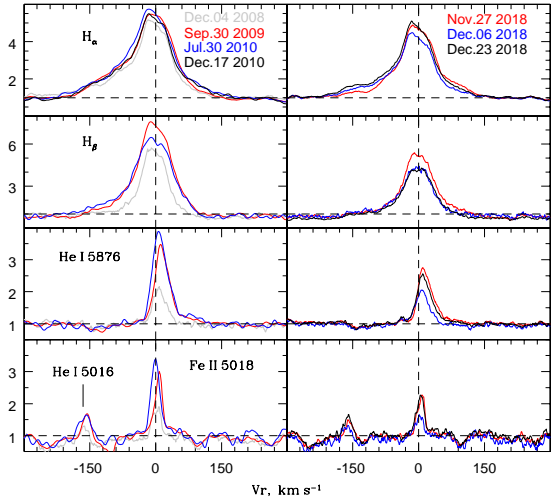}
 \end{center}
  \caption{ The profiles of some emission lines in the 2008-2010 (left 
column) and 2018 (right column) spectra of ZZ~Tau\,A+B. According to
Table~\ref{tab:spectra}, the spectra from the right column corresponding to
dates 2018-Nov-27, 2018-Dec-07 and 2018-Dec-24 are referred to in the text as
the abbreviation UVES-1, UVES-2 and UVES-3, respectively. The profiles are
plotted in stellar rest frame assuming $V_{\rm r}\approx 17.3$ km~s$^{-1}$
for the left panel and 15.4 km~s$^{-1}$ for the right one.} 
 \label{fig:HHeFe2-Em-lines}
\end{figure}

\begin{figure}
 \begin{center}
\includegraphics[scale=0.5]{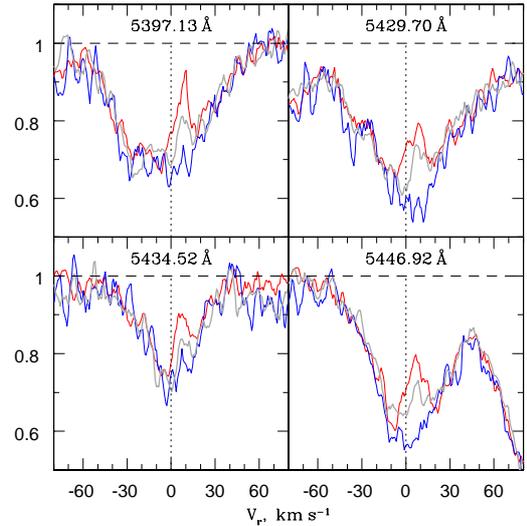}
 \end{center}
  \caption{The profiles of some \ion{Fe}{i} a$^5$F - z$^5$D$^{\rm o}$
multiplet lines in the 2018 November 27 (UVES-1, red), December 06
(UVES-2, blue) and December 23 (UVES-3, gray) spectra of ZZ~Tau\,A+B. 
The radial velocity along the abscissa axis refers to the rest frame of the star. }
 \label{fig:FeI-profs}
\end{figure}

  We have identified a number of emission lines in the ZZ~Tau\,A+B spectra -- see
Figs.\ref{fig:muse-emiss-lines}--\ref{fig:FeI-profs}. Among the most prominent are 
the Balmer lines of hydrogen (from H$\alpha$ to at least H$\delta$), \ion{He}{i}
(triplet $\lambda\lambda$ 4471.5, 5875.6 and singlet $\lambda\lambda$
4921.9, 5015.7, 6678.2~\AA{}), \ion{Ca}{ii} (H, K, IR triplet),
\ion{Mg}{i} $\lambda\lambda$5167.3, 5172.7~\AA{} lines as well as the strongest
lines of \ion{Fe}{i} $(a^5{\rm F}-z^5{\rm D}^{\rm o}$, $a^3{\rm F} - z^3{\rm
F}^{\rm o},$ $a^3{\rm F} - z^3{\rm D}^{\rm o})$ and \ion{Fe}{ii} $(a^6{\rm}S
- z^6{\rm P}^{\rm o},$ $a^4{\rm G} -z^4{\rm F}^{\rm o})$ multiplets.  All the
iron lines have $\log (gf)>-1.9$, and an excitation energy of the low level is $E_{\rm
i}\lesssim 3.3$~eV.

%
%


 As can be seen from Fig.\ref{fig:MUSE-vs-CMO} and
\ref{fig:muse-emiss-lines} the emission lines are present in the spectra of
both components.  In particular, the H$\alpha$ equivalent width is $22\pm
1$ and $13 \pm 0.5$~\AA{} in the {MUSE} spectra of ZZ~Tau~A and B, and that
of H$\beta$ is $\approx 13$ and 11~\AA{}, respectively.  {\it Observed}
secondary to primary H$\alpha$ flux ratio in the {MUSE} spectra is 0.30 and
H$\beta$ flux ratio is 0.71.  It means that the relative contribution of
ZZ~Tau~B to the H$\alpha$ and H$\beta$ fluxes in the total ZZ~Tau~A+B
spectra are $\sim 20$ and 40\,\%{}, respectively.

%
%

   In Fig.~\ref{fig:HHeFe2-Em-lines} we compared the profiles of some
\ion{H}{i}, \ion{He}{i} and \ion{Fe}{ii} emission lines from the high
resolution spectra of ZZ~Tau~A+B observed in 2008-2010 (left panel) and 2018
November-December. It can be seen that the profiles and EWs of these lines
are variable. In particular, the EW of $H\alpha$ varied between 8.8 to
17~\AA{} in the 2008-2010 spectra, and between 8.5 and 9.4~\AA{} in the 2018
{UVES} spectra. But the profiles of the same lines in both panels have not
changed qualitatively, so it seems reasonable to compare the short term line
profile variability with periodical photometric variability.

  According to Table~\ref{tab:spectra} the phase differences between the
moments of obtaining the UVES-2 and UVES-3 spectra relative to that of the
UVES-1 one are $\approx 2.39 \pm 0.005$ and $6.42\pm 0.01$, respectively,
for the rotation period $P_{\rm rot}=4{\fd}171 \pm 0.002$.  If the observed
line variability is related to the rotational modulation, then one can
expect that the line profiles should be nearly identical in the UVES-2 and
UVES-3 spectra and differ from those in the UVES-1 spectrum.  But as can be
seen from the right panel of Fig.~\ref{fig:HHeFe2-Em-lines} this is true
only for the H$\beta$ line, and for all other lines the situation is quite
the opposite: the profiles in the UVES-2 and UVES-3 spectra differ
significantly, but are more or less similar in the UVES-1 and UVES-2
spectra.  This can be seen even more clearly in the case of the \ion{Fe}{i}
a$^5$F - z$^5$D$^{\rm o}$ multiplet lines shown in Fig.\ref{fig:FeI-profs}. 
We will try to explain this discrepancy (the {\lq}phase problem{\rq}) a bit
later.

%
%

\begin{figure}
 \begin{center}
\includegraphics[scale=0.4]{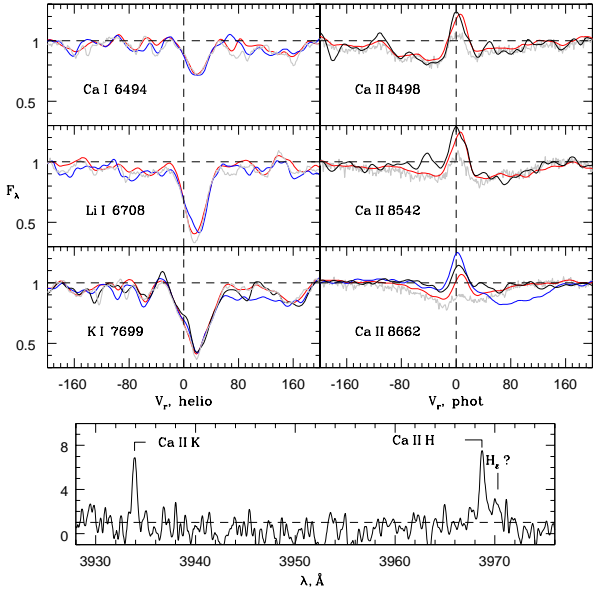}
 \end{center}
  \caption{The profiles of some lines in the 2008 Dec.04 (grey), 2009 Sep.30 (red),
2010 Jul.30 (blue) and 2010 Dec.17 (black) spectra of ZZ~Tau\,A+B.  Left
column -- the absorption lines in heliocentric frame, right column -- the
\ion{Ca}{ii} infrared triplet lines in the stellar rest frame assuming
$V_{\rm r}\approx 17$~km\,s$^{-1}.$ A portion of ZZ~Tau spectrum averaged
over all the 2009-2010 observations and additionally smoothed with a running average over 10 points is
shown in the middle bottom panel.  }
 \label{fig:abs-and-CaII-profs}
\end{figure}

   The \ion{He}{i} and \ion{Fe}{ii} emission lines in
Fig.~\ref{fig:HHeFe2-Em-lines} as well as the emission components of
\ion{Fe}{i} lines in Fig.~\ref{fig:FeI-profs} are redshifted.  Using our
orbital parameters one can find that at the middle epoch of the UVES spectra
observations (the beginning of December 2018) the true anomaly $v$ was
$\approx 181 \degr.$ Then, according to Eq.(\ref{eq:delta-Vr-1}), the
difference of the component's radial velocities $V_{\rm r}^{\rm B} - V_{\rm
r}^{\rm A}$ was $\approx -1.6$~km~s$^{-1}$ at that moment in the case of the
{\lq}mirror{\rq} orbit.  As can be seen from the figures, the maximum of
\ion{He}{i}, \ion{Fe}{i} and \ion{Fe}{ii} line profiles in the UVES spectra
are redshifted greater -- from 5 to 10~km~s$^{-1}.$ Probably it means that
they are really redshifted in the spectra of the primary, but it is not so
obvious in the case of ZZ~Tau~B.  The situation with the emission components
of the infrared \ion{Ca}{ii} triplet lines in the ZZ~Tau~A+B spectra (see
Fig.\ref{fig:abs-and-CaII-profs}), is similar: their profiles are also
variable and slightly redshifted.  Unfortunately, the spectral resolution of
the {MUSE} spectrum is too low to detect the redshift of emission lines for
each component which is of the order of several km~s$^{-1}.$

  The average H$\alpha$ line flux in the three {UVES} ZZ~Tau~A+B spectra is
$\approx 1.3\times 10^{-13}$~erg~s$^{-1}$~cm$^{-2}.$ If we suppose that this
value stayed constant during the {MUSE} spectral observations (one year later)
and use the component's flux ratio mentioned above, we can find the observed as
well as de-reddened \citep{Cardelli+1989} H$\alpha$ line flux $F_{\rm
H\alpha}$ of each component of the binary, and their respective luminosities
$L_{\rm H\alpha}=4\pi d^2 F_{\rm H\alpha},$ where $d=140$~pc is the distance
to ZZ~Tau.  Thus, we found $L^{\rm A}_{\rm H\alpha}\approx 7.3\times
10^{29}$~erg~s$^{-1}$ and $L^{\rm B}_{\rm H\alpha}\approx 1.8\times
10^{29}$~erg~s$^{-1}.$

  To estimate the accretion luminosity $L^{\rm A}_{acc}$ of ZZ~Tau~A we used the
statistical relation 
\begin{equation}
\log \left( {L_{acc} \over L_\odot} \right) = 
1.13\, \log \left( {L_{\rm H\alpha} \over L_\odot} \right) + 1.74
 \label{eq:alcala}
\end{equation}
of \citet{Alcala-2017} and found that $L^{\rm A}_{\rm acc}\approx 1.3\times
10^{31}$~erg~s$^{-1}$. Then we estimated the ZZ~Tau~A's accretion rate from
Eq.1 of \citet{Alcala-2017} and found $\dot M^{\rm A}_{\rm acc} \sim 7\times
10^{-10}$~M$_\odot$~yr$^{-1}.$ Applying the same procedure to ZZ~Tau~B leads
to a four time less $\dot M^{\rm B}_{\rm acc}$. 

   As follows from Fig.~B.1 of \citet{Alcala-2017} both components of the
binary are located in the region of the $\log L_{acc}$ vs $T_{\rm eff}$
diagram where the chromospheric contribution to the line emission is
negligible.  Thus, it seems reasonable to conclude that accretion rather
than cromospheric activity is responsible for the line emission observed in
both components of ZZ~Tau. Recall also that \citet{Espaillat-2012} concluded
that at least one component of the binary has an optically thick accretion
disc.

  Our value of $\dot M^{\rm A}_{\rm acc}+\dot M^{\rm B}_{\rm acc}$ is
between the accretion rates found by \citet{Cieza-2012} and
\citet{Valenti-1993} from spatially unresolved ZZ~Tau~A+B's spectra.  Note
however that Eq.~\ref{eq:alcala} is statistical in its nature, so our
estimate of $\dot M_{\rm acc}$ is valid up to a factor of 2-3.  A more
accurate estimation can be found via a respective analysis of a series of
high resolution spectra of each component of the binary \citep{Dodin-13b}. 
Nevertheless, if our estimate is correct to the order of magnitude, then the
time for the circumprimary disc to exhaust its matter is $M_{\rm disc}/\dot
M^{\rm A}_{\rm acc} < 3\times 10^5$~yr, i.e. significantly less than the
age of the binary.  So, one can expect that there is a matter flow from the
CB disc at least to the disc of the primary.

%
%

  Let's get back to the {\lq}phase problem{\rq} now, assuming that the observed
photometric periodicity results from the rotational modulation of the accretion
hot spot {\it continuum} emission. We have mentioned in Sect.~\ref{subsect:photometry}
(see Fig.~\ref{fig:period}) that the $4{\fd}17$ phase curve is clearly seen in the
$V,$ $R,$ $I$ bands, becomes blurred in the $B$ band and that the periodicity is not
seen in the $U$ band.  At the same time the relative contribution of ZZ~Tau~B to the
total flux of ZZ~Tau~A+B begins to increase with wavelength decreasing at
$\lambda \lesssim 0.6$~\micron{} (see Fig.~\ref{fig:MUSE-vs-CMO}), so we
believe that the emission of ZZ~Tau~B is much stronger in the $U$ band than that of
the primary.  One can explain these facts if the accretion hot spot
responsible for the observed photometric periodicity is located on ZZ~Tau~A.

  On the other hand, we could not reliably detect the periodicity in the $U$
band, where presumably ZZ~Tau~B is the main source of emission.\footnote{More precisely we found the period $P=2{\fd}237,$ but its false alarm probability is $\approx 10$~\%{}, i.e. large enough.  Moreover this value of $P$ does non solve the {\lq}phase problem{\rq}.}

It could be due e.g. to the angle between the rotational and magnetic axes
of the secondary being $\lesssim 30\degr$ and the accretion flow being
divided into several unstable {\lq}tongues{\rq} that produce chaotic hot
spots on the stellar surface resulting in an irregular light curve -- see
\citet{Romanova-2021} and references therein.  If so, one can explain the
{\lq}phase problem{\rq} in the following way.

  As was noted, the contribution of the secondary to the total radiation
of ZZ~Tau~A+B is approximately the same in line and continuum emission and
varies from 20 to 40~\%{} in the range from 0.5 to 0.9~\micron. This is
enough to noticeably affect the shape of the lines in the total spectrum of
the binary, and to increase the noise level of the photometric phase curve,
and the effect is the greater, the greater relative flux contribution of the companion. And where
the radiation of ZZ~Tau~B dominates (e.g. in the $U$ band) photometric
periodicity disappears.

  It was mentioned in Sect.\ref{subsect:photometry} that the binary
became brighter in all the bands in 2021 than in 2020, and to a greater extent in
the short-wavelength region (Fig.\,\ref{fig:lcCMO}).\footnote{The accuracy 
of our speckle interferometric measurements is not enough to determine, which
component of the binary has become brighter: as we noted in 
Sect.~\ref{sect:observation} the difference of component's relative brightness
in 2020 and 2021 observational seasons at $\lambda \approx 0.9$~\micron{} is
less than the errors of measurements.}
The amplitude of photometric variability in the historical light curve is
also significantly larger in the $B$ than $V$ band (Fig.\ref{fig:lcHist}). 
This suggests that the secondary rather than the primary is responsible for
the long term photometric variability of ZZ~Tau. The chaotic accretion
regime \citep{Romanova-2008} may be responsible for an increased photometric
activity of ZZ~Tau~B. Note in this regard that there is no obvious
relation between the historical light curve and relative position of the
binary components.

  It's especially worth mentioning that we did not find any gas outflow signs in the
spectra of ZZ~Tau~A+B: there are neither forbidden lines nor clear excess emission
in blue wings of permitted lines which is in agreement with the observations of other
authors \citep{Valenti-1993, Kenyon-1998, Folha-2001,
Herzeg-Hillebrandt-2014}.

 At the end of this section we would like to draw attention to the
following.  As was demonstrated above, the variable redshifted emission component
is present in the red wing of \ion{Fe}{i} lines, no matter which component
of the binary they are related to.  It well can be that other absorption
lines also have similar (but not so strong) variable redshifted emission, as
a result of which the centroid of the lines will be shifted, and this
ultimately leads to an error in determining the radial velocity of the
companion or the primary \citep{Dodin-13b}. The discrepancies in
the calculated and observed radial velocities of ZZ~Tau's components (see
Table~\ref{tab:Prato}) may be partly due to this effect.

%
\section{Circumstellar environment of the binary}
 \label{sect:environment}

%
\begin{figure*}
 \begin{center}
\includegraphics[scale=0.4]{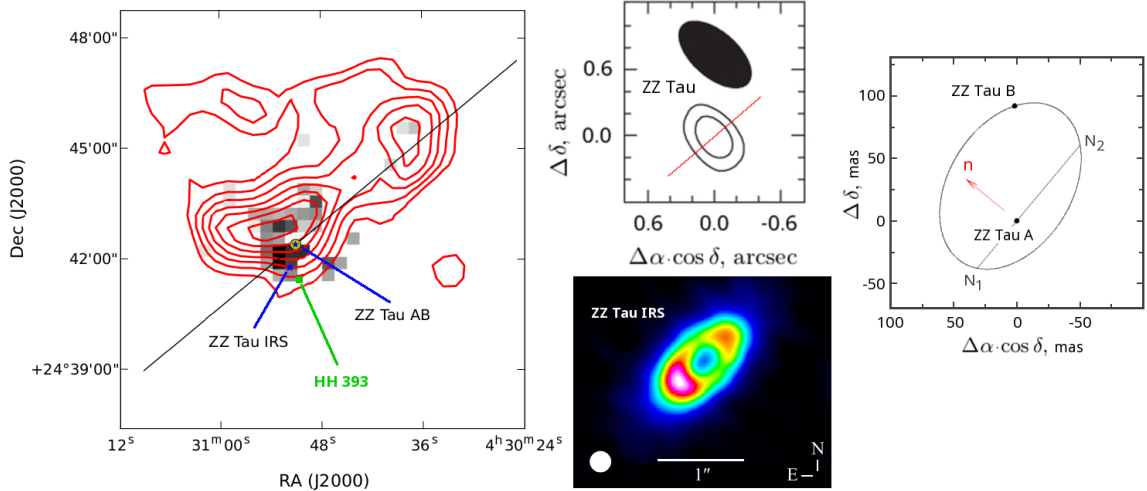}
  \caption{Left panel -- a monopolar redshifted CO molecular outflow of
ZZ~Tau (red isophotes) and its axis (black line) according to
\citet{Narayanan-2012}.  The positions of ZZ~Tau~IRS and HH393 are also shown. 
Middle panels: top -- the isophotes of ZZ~Tau's CB disc (solid lines) and the
corresponding beam size \citep{Akeson-2019}; bottom -- the protoplanetary disc
of ZZ~Tau~IRS \citep{Hashimoto-2021}.  The red line is the direction of the
CO outflow.  Right panel -- the relative orbit of ZZ~Tau binary shown for
comparison.  N$_1$-N$_2$ is a node line, and $n$ is the projection of
orbital momentum vector to the sky.  See text for details.
 \label{fig:morphology}
}
 \end{center}
\end{figure*}
%

  As was noted above there are no indications of matter outflow in the optical
and near IR spectra of ZZ~Tau\,A+B, but there is a monopolar redshifted
${}^{12}$CO molecular outflow in the vicinity of the binary
\citep{Heyer-1987}. In the left panel of Fig.~\ref{fig:morphology} we
reproduce a portion of Fig.27 of \citet{Narayanan-2012}, who analysed the
velocity field of the outflow and concluded that the position angle of its axis
is PA$_{\rm CO} \approx 130\degr$ -- see the black line in the panel. 
\citet{Narayanan-2012} noted that it is not possible to decide whether ZZ~Tau
or ZZ~Tau~IRS is the source of the outflow, relying only on its morphology. 

To solve the problem one can try to use information about
orientation of protoplanetary discs of these objects assuming that the outflow
axis is perpendicular to the disc plane.  In this case the axis of the
CO-outflow should be oriented along the minor axis of ALMA image of
(presumably round shape) respective disc.  Unfortunately, the continuum CB
disc of ZZ~Tau measured by \citet{Akeson-2019} appears to be smaller than
the beam size and therefore unresolved -- see the upper middle panel of
Fig.~\ref{fig:morphology}.  By contrast, the circumstellar disc of ZZ~Tau~IRS is
fully resolved (see the middle bottom panel of Fig.~\ref{fig:morphology})
and the position angle of its minor axis is $45\degr$ \citep{Hashimoto-2021}. 
It means that the minor axis of ZZ~Tau~IRS disc is almost perpendicular to
the axis of CO outflow and pointing almost to the Herbig-Haro object HH393. 
On this basis, we believe that ZZ~Tau binary is the source of the
CO-outflow, but ZZ~Tau~IRS is the source of HH393 outflow. 

   Additional more sensitive and with better angular resolution {ALMA}
observations of ZZ~Tau's CB disc are needed to determine its inclination
$i_{\rm d}$ and the position angle of the disc's ascending node $\Omega_{\rm
d}.$ It will then be possible to determine the inclination $\Theta$ of
ZZ~Tau~A+B's orbital plane to the CB disc of the binary as follows. The
projection of the angular momentum unit length vectors of the disc {\bf
n}$_{\rm d}$ and orbital motion {\bf n}$_{\rm o}$ in our coordinate system
(see the footnote to Sect.~\ref{subsect:orbit}) is: $\{ \cos i_{\rm j}\,
\cos \Omega_{\rm j}; \, \sin i_{\rm j}\, \sin \Omega_{\rm j}; \, \cos i_{\rm
j} \},$ where $j = {\rm d}$ and ${\rm o},$ respectively. Then, the angle
$\Theta$ between the {\bf n}$_{\rm d}$ and {\bf n}$_{\rm o}$ vectors can be
found from their scalar product:
\begin{equation}
\cos \Theta = 
\cos i_{\rm d}\, \cos i_{\rm o} + 
\sin i_{\rm d} \, \sin i_{\rm o} \cos \left( \Omega_{\rm d}-\Omega_{\rm o} \right)
\, .
 \label{eq:disc-orbit-inclin}
\end{equation}

  As far as we have proposed that the observed periodical brightness variations of
ZZ~Tau in the $BVRI$ bands are related to axial rotation of the primary,
then one can estimate the inclination $i_*$ of its axis to the line of
sight from the relation:
\begin{equation}
\sin i_* = { P_{\rm rot} \times v\sin \,i \over 2\pi R } \, .
 \label{eq:sini}
\end{equation} 
According to L.~Prato (private communication) the projected rotational
velocities $v\sin i$ of ZZ~Tau~A and ZZ~Tau~B are $11 \pm 1$ and $17\pm 1$~
km~s$^{-1},$ respectively.\footnote{ \cite{IGRINS-2021} and \cite{Nofi-2021}
reported $v\sin i=20.6 \pm 2$ and $18.4 \pm 2.4$~km~s$^{-1},$ respectively
for ZZ~Tau~A+B.} It means that the axis of ZZ~Tau~A is inclined to the line of sight at $\approx 31 \degr \pm 4 \degr.$

 The absence of detectable blueshifted CO emission near ZZ~Tau (see the left
panel of Fig.\ref{fig:morphology}) implies a small amount of cold gas (the
remnants of the parent protostellar cloud) in front of the binary.  The
polarization of ZZ~Tau's radiation (see Table~\ref{tab:polarim}) is also
small $(p_{\rm I} \approx 0.27$~\%{}) and stable.  It can be explained by
interstellar dichroic absorption.  Taking into account the Serkowski law
determined for Elias~9 \citep{Whittet-1992}, which is $0\fdg317$ away from
ZZ~Tau, $p_{\rm V}$ for ZZ~Tau should be $\approx 0.12~\%{},$ that
corresponds to the interstellar $A_{\rm V}\gtrsim 0.04$.  Thus, we concluded
that a relatively large extinction in the direction to the components
of the binary $(A_{\rm V}^{\rm A}\approx 1.5$ and $A_{\rm V}^{\rm B} \approx
1.2$ according to \citet{Schaefer-2014}) is related to the matter in the
immediate vicinity of these stars. Recall in this connection that the
matter flow from the CB disc to the disc of ZZ~Tau~A can still continue.


\section{Summary}
\label{sect:summary}

   The results of our investigation of ZZ~Tau can be summarized as follows.
\begin{enumerate}
\item  We clarified the orbital parameters of the binary using published and our
new astrometric observations. It appeared that our parameters differ from
those of \citet{Schaefer-2014} by $< 1\,\sigma$, but have $1.5-2$ times
better accuracy.  
\item According to our calculations, the parameters of the binary are: 
the orbital period of the binary $P=46.8 \pm 0.8$~yr, semimajor axis
$a=12.35 \pm 0.29$~au, eccentricity $e= 0.58 \pm 0.02,$ inclination of the orbit
$i=123{\fdg}8 \pm 1{\fdg}0,$ and total mass of the binary $0.86 \pm
0.09$~M$_\odot.$ In particular, this means that the minimal distance between
the components is $\approx 5.2$~au.
\item It is not possible to determine the direction of the ascending node
from a visual orbit alone, so the two solutions with $(\omega,\Omega)$ and
$(\omega+180\degr{},\Omega+180\degr{})$ are equivalent from a visual binary
perspective. We used the radial velocities to break the degeneracy between
these two solutions and determine the direction of the ascending node.
But it is a preliminary conclusion, because it appeared that a variable
emission component can present in the red wing of (at least some)
absorption lines, thus introducing an error to the companion's radial
velocity measurements.  High resolution spatially resolved spectra of
ZZ~Tau's components are required to take this effect into account.
\item  The radiation of ZZ~Tau~B dominates shortward of $\lambda \approx
0.4$~\micron{} in the SED of ZZ~Tau~A+B, despite the fact that $T^{\rm
B}_{\rm eff} < T^{\rm A}_{\rm eff}.$  In our opinion, this means that the
extinction in the direction to the primary is larger then to the secondary
presumably due to an excess amount of circumstellar material in the vicinity of
ZZ~Tau~A.
\item We found that the rotation axis of the primary is inclined to the line of sight by $\approx 31 \degr \pm 4 \degr.$
\item  Using the isochrones of \citet{Baraffe+2015}, we conclude that 
the age of both ZZ~Tau's components is between 0.5 and $2$~Myr.
\item  The emission lines are observed in the spectra of both ZZ~Tau's
components, e.g. the equivalent width of H$\alpha$ line in the {MUSE}
spectra of ZZ~Tau~A and B is $\approx 22$ and 13~\AA{}, respectively.  We
concluded that the observed line as well as the continuum excess emission in
the spectra of both components is due to accretion of disc matter. 
According to our estimate, the accretion rate to the primary is $\dot M^{\rm
A}_{\rm acc} \sim 7\times 10^{-10}$~M$_\odot$~yr$^{-1}$ and $\approx 4$
times less in the case of ZZ~Tau~B.
\item The periodical light variations of ZZ~Tau~A+B with a period of $P=
4{\fd}1710\pm 0{\fd}0022$ were observed in the $BVRI$ bands that is close to
the period of $4{\fd}1609$ reported by \citet{Rebull-2020}.  It follows from
our data that the period $1{\fd}311$ \citep{Rodriguez-2017} is probably the
one day alias of the real one.  We believe that the periodicity is related
to axial rotation of an accretion (hot) spot on the surface of the primary.
\item We did not observe a $4{\fd}17$ periodicity in the $U$ band or in the 
variations of emission line profiles presumably due to a significant
contribution of the ZZ~Tau~B emission.
\item There is no obvious correlation between long-term photometric
variability of ZZ~Tau and orbital position of the binary components.  We
believe that ZZ~Tau~B is responsible for the large amplitude photometric
variability of the binary observed in the first half of the last century.
\item Judging by the orientation of ZZ~Tau~IRS's disc, it seems 
much more likely that ZZ~Tau~A+B is the source of CO-molecular outflow but
ZZ~Tau~IRS is the source of Herbig-Haro object HH393.
\end{enumerate}

  Apparently our estimation of the binary orbital parameters is reliable
enough.  It is necessary to obtain high resolution optical spectra of each
component at different phases of rotation period to get more reliable
estimates of the component's parameters (effective temperatures,
extinction(s), luminosities, radii, and accretion rates).  We also propose
to include ZZ~Tau in the target list of spectral UV observations with
{WSO-UV} space mission \citep{Boyarchuk-2016}.  Additional more sensitive
and with better angular resolution {ALMA} observations of ZZ~Tau are highly
desirable to get important information on the parameters of its CB disc.  We
hope that these new observations will help to clarify whether the
extinction in the direction of the binary components is different and to
understand the reason for the increased photometric activity of ZZ~Tau in
the first half of the last century.


%
\section*{Acknowledgements}

We thank the staff of the Caucasian Mountain Observatory headed by N.~Shatsky for the help with observations as well as Regina von Berlepsch and Matthias Steffen for sending us a copy of the \citet{Himpel-1944} paper. 
We also thank N.N.~Samus and L.M.~Rebull for useful discussions and especially Dr.  L.~Prato who has sent us the unpublished results of $V_r$ and $v\sin i$ measurements for the ZZ~Tau components. We give our special thanks to the anonymous reviewer for the comments made, which allowed us to eliminate shortcomings and inaccuracies presented in the original version of the paper. We acknowledge with thanks the variable star observations from the AAVSO International Database contributed by observers worldwide and used in this
research.  This research has made use of the SIMBAD database, operated at
CDS, Strasbourg, France.  The study of AD (observations, data reduction,
interpretation) and SL (interpretation) and BS (polarimetric observations and data reduction) was conducted under the financial support of the Russian Science Foundation 17-12-01241.

The acquisition of scientific equipment used in this study was partially supported by 
the M.~V.~Lomonosov Moscow State University Program of Development.

			
\section*{Data availability}

The photometric data used in this article are available in online
supplementary material. Other data used in this article will be
shared on reasonable request to the corresponding author.

\bibliographystyle{mnras}
\bibliography{lamzin2021}

\bsp	

\label{lastpage}
\end{document}